\documentclass{llncs}
\usepackage{amssymb} % for drawing double-barred R, Z, C, etc.
\usepackage{epsfig}

\begin{document}
\pagestyle{plain}  
\bibliographystyle{splncs}

\title{A Constructive Generalization\\of Nash Equilibrium\\ for Better Payoffs and Stability}

\author{Xiaofei Huang}

\institute{eGain Communications\\Mountain View, CA 94043, U.S.A. \\
\email{huangxiaofei@ieee.org}}

\maketitle
%

%%%%%%%%%%%%%%%%%%%%%%%%%%%%%%%%%%%%%%%%%%%%%%%%%%%%%%%%%%%%%%%%%%
\begin{abstract}
%%%%%%%%%%%%%%%%%%%%%%%%%%%%%%%%%%%%%%%%%%%%%%%%%%%%%%%%%%%%%%%%%%
In a society of completely selfish individuals where everybody is only interested in maximizing his own payoff, 
	does any equilibrium exist for the society?
John Nash proved more than 50 years ago that an equilibrium always exists
	such that nobody would benefit from unilaterally changing his strategy.
Nash Equilibrium is a central concept in game theory,
	which offers a mathematical foundation for social science and economy.
However, it is important from both a theoretical and a practical point of view 
	to understand game playing where individuals are less selfish.
This paper offers a constructive generalization of Nash equilibrium
	to study $n$-person games where the selfishness of individuals can be defined at any level, 
	including the extreme of complete selfishness.
The generalization is constructive since it offers a protocol for individuals in a society to reach an equilibrium.
Most importantly, this paper presents experimental results and theoretical investigation 
	to show that the individuals in a society can reduce their selfishness level together
	to reach a new equilibrium where they can have better payoffs
	and the society is more stable at the same time.
This study suggests that, for the benefit of everyone in a society (including the financial market), 
	the pursuit of maximal payoff by each individual should be controlled at some level
	either by voluntary good citizenship or by imposed regulations.
\end{abstract}

%------------------------------------------------------------
\section{Introduction}
%------------------------------------------------------------
John Nash has proved in 1950~\cite{nash50} using Kakutani fixed point theorem that any $n$-player normal-form game~\cite{GameTheoryLuce}
	has at least one equilibrium.
In the game, each player has only a finite number of actions to take and 
	takes one strategy at action playing.
If a player takes one of the actions in a deterministic way,
	it is called a pure strategy.
Otherwise, if a player takes anyone of the actions following some probability distribution defined on the actions,
	it is called a mixed strategy.
At a Nash equilibrium,
	each player has chosen a strategy (pure or mixed)
	and no player can benefit by unilaterally changing his or her strategy while the other players keep theirs unchanged. 

Nash Equilibrium is arguably the most important concept in game theory,
	which has significant impacts on many other fields like social science, economy, and computer science.
It is an important theory for understanding a common scenario in game playing.

In a Nash equilibrium, each player's strategy is completely selfish
	because the player is only interested in maximizing his own payoff. 
Only the best action(s) to each player is accepted by the player,
	sub-optimal actions are not considered at all.
The best action is defined as the one with the highest payoff.
As a consequence of the selfishness, even if the payoff of a sub-optimal action is slightly less than the best one,
	the probability of picking this sub-optimal action by the player is still zero.

However, many cultures teach people to be less selfish in a society.
Also, the scenario of less-selfish players may be closer to reality, 
	such as individuals in human societies or animal kingdoms. 
Our conventional wisdom tells us
	that if each of us gives away a bit more in favor of others,
	we could end up with more gains as return.
That is, reduced selfishness leads to better payoffs for the individuals in a society.
For instance, if we, as drivers, 
	respect other drivers sharing the same road and 
	give considerations for each other either voluntarily and/or by following traffic laws,
	then each of us will end up with a faster, safer drive to his/her destination 
	than the case when everyone is only interested in maximizing his own speed to his destination.

This paper presents both experimental results and theoretical investigation 
	to show that, if the individuals in a society reduce their selfishness 
	by simply accepting sub-optimal actions in some degree, 
	a new equilibrium can be reached where better payoffs and social stability are obtained at the same time.

The first key observation of this paper is that, reducing selfishness can improve payoffs.
When completely selfish players at a Nash equilibrium reduce their selfishness, 
	they will shift to a new equilibrium with payoffs possibly better than the original one.
The observations will range from the classic prisoner's dilemma, 
	a hard game used in other game theory literatures, to computer generated games with hundreds to thousands of players.
It verifies the conventional wisdom that reducing selfishness could lead to better payoffs for everyone.

The second key observation is that, reducing selfishness 
	can also improve social stability.
A society of completely selfish individuals can be very sensitive to perturbations,
	the accuracy at representing individuals' utility functions,
	and communication errors among the individuals in the society.
The smallest change in utility function or the slightest communication error could knock the individuals
	out of their existing equilibrium.
Furthermore, a society of completely selfish individuals can have an enormous number of equilibria.
The number may increase exponentially with the population of the society,
The society could end up with one Nash equilibrium or another,
	depending on the initial conditions and sensitive to perturbations.
If the individuals reduce their selfishness together, 
	they can reduce their sensitivity to perturbations, inaccuracy in utility functions, and communication errors.
At the same time, the number of equilibria tends to drop significantly 
	so that the outcome of the society can be more predictable.
When the selfishness is below a certain level, 
	the society tends to have only one equilibrium and converges to it with any initial conditions.
	
In particular, this paper gives a mathematical model for describing selfishness.
The level of selfishness is controlled by one parameter of the model 
	to cover the spectrum ranging from complete selfishness to complete selfishlessness.
With the parameterized selfishness model, 
	this paper offers a generalization of Nash equilibrium 
	together with a proof of the existence of an equilibrium given any selfishness level using a fixed point theorem.
It is a generalization because this paper offers a proof to show that 
	a generalized equilibrium at the particular case of completely selfish players
	falls back to a Nash equilibrium.
In other words, the definition of Nash equilibrium is a special case of the generalized one.
It is important to note that the generalization 
	is constructive because it defines a protocol for the players in a game 
	to interact with each other so that an equilibrium can be reached with any selfishness level.

%The generalization of Nash equilibrium described in this paper is along 
%	the line of the player's selfishness level.
%It also offers experimental results and a theoretical investigation 
%	to show that the individuals in a society can reduce their selfishness level together
%	so that better payoffs and stability can be obtained.
%It sheds new insights from a mathematical perspective for building a good and stable economy or society,
%	especially at the time of the global economic crisis.

%-----------------------------------------------------------------------------------------
\section{A Constructive Generalization}	
%-----------------------------------------------------------------------------------------

An $n$-player normal form game is defined as:
\begin{itemize}
\item $n$ players $1,2,\ldots, n$;
\item Each player $i$ has a finite set of strategies $S_i = \{s_{i1}, s_{i2}, \ldots, s_{im_i}\}$. 
Strategies are also called actions.
The Cartesian products of $S_i$, $S = S_1 \times S_2 \times \cdots \times S_n$, 
	is called the set of the pure strategy profiles (the set of action tuples). 
\item Each player has a utility function defined as a real value function $u_i(x)$ defined on the set of the pure strategy profiles $S$,
	i.e., $u_i(x): S \rightarrow \mathbb{R}$ (a mapping from each action tuple to a real value).
\end{itemize}

If player $i$ takes one of the actions from $S_i$ in a deterministic way,
	it is called a pure strategy.
Otherwise, if the player takes any of the actions following some probability distribution $p_i$ defined on the action set $S_i$,
	it is called a mixed strategy.
That is, for each action $x_i \in S_i$, the player $i$ takes this action with a probability $p_i(x_i)$.
A set of (mixed) strategies $\{p_1, p_2, \ldots, p_n \}$, one for each player, is called a (mixed) strategy profile $p$.

Assume that the $n$ players take a strategy profile $p$. 
Then the payoff of player $i$ is defined as 
\[ u_i (p) = \sum_{x \in S} u_i (x) \prod_{j} p_j(x_j), \quad \mbox{for $i=1,2,\ldots, n$}  \ . \]
The objective of each player is to maximize his payoff.

A strategy profile excluding the one for player $i$ is denoted as $p_{-i}$.
A strategy profile $p^{*}$ is a Nash equilibrium if for all $i$ and for all $p_i$,
\[ u_i (p_i, p^{*}_{-i}) \le u_i (p^{*}_i, p^{*}_{-i}) \ . \]
That is, no unilateral deviation in strategy by any player gives higher payoff
	for that player.
Nash's 1950 PNAS paper proves the existence for an equilibrium for any finite $n$-player game using Kakutani's fixed point theorem.

In the following discussions, without loss of generality,
	we assume all utility functions are of positive function values, i.e.,
	$u_i(x) > 0$, for any $x \in S$.

If player $i$ takes an action $x_i \in S_i$ in response to other players strategies $p_{-i}$,
	the payoff is $ u_i (x_i, p_{-i})$.
The optimal action $x^{*}_i$ for the player $i$ is defined as the one with the highest payoff, i.e.,
\[ u_i (x^{*}_i, p_{-i}) = \max_{x_i \in S_i} u_i (x_i, p_{-i}) \ . \]
Obviously,
\[ u_i (p_i, p_{-i}) = \sum_{x_i \in S_i} p_i(x_i) u_i (x_i, p_{-i}) \ . \]

One of the important properties of a Nash equilibrium $p^{*}$ is that 
	only the optimal action(s) has non-zero probability, i.e., 
	if $p^{*}_i(x_i) > 0$, then $x_i$ must be the optimal action for the player $i$.
In other words, player $i$ is completely selfish because he only accepts the optimal action for himself.

Assume that, at a time instance $t$, the strategy profile excluding the one for player $i$ is $p_{-i}(t)$,
	the action payoff for player $i$ is $u_i(x_i, p_{-i}(t))$, for $x_i \in S_i$.
Based on the above observation,
    we can define a mathematical model
    to formulate the construction of the next time strategy $p_i(x_i, t+1)$ for player $i$ 
    based on his action payoff $u_i(x_i, p_{-i}(t))$ at the current time $t$.
Specifically, we can construct $p_i(x_i, t+1)$ such that it is proportional to $u_i(x_i, p_{-i}(t))$, e.g.,
\[ p_i (x_i, t+1) \propto \left(u_i (x_i, p_{-i}(t))\right)^{\alpha}, \quad \mbox{for $x_i \in S_i$}  \ , \]
where $\alpha$ is a parameter of a non-negative value.
Since $p_i(x_i, t+1)$ should be normalized as a probability, the above formula can be rewritten as
\begin{equation}
p_i (x_i, t+1) = \frac{\left(u_i (x_i, p_{-i}(t))\right)^{\alpha}}{ \sum_{x_i \in S_i} \left(u_i (x_i, p_{-i}(t))\right)^{\alpha}},  \quad \mbox{for $i=1,2,\ldots, n$} \ . 
\label{construct-new-strategy}
\end{equation}

In (\ref{construct-new-strategy}), 
	when $\alpha \rightarrow \infty$, the best action has a non-zero probability while others have probability zero.
That is, the player only accepts the best action, the one with the highest payoff $u_i(x_i, p_{-i})$.
It is exactly same as the case of Nash equilibrium described before.

If the value of $\alpha$ is reduced from the above extreme case, 
	the player $i$ starts to accept sub-optimal actions by assigning non-zero probability to them.
The degree of the acceptance increases with further decrease of $\alpha$.
At another extreme case, when $\alpha \rightarrow 0$,
	each action is assigned with the same probability 
	and the player has no preference on any one of the actions.
All of the actions are treated equally and they are sampled uniformly.
In this case, the player is completely selfishless.
In summary, the parameter $\alpha$ describes the selfishness level of player $i$. 
It covers the spectrum ranging from complete selfishness ($\alpha \rightarrow \infty$) to complete selfishlessness ($\alpha = 0$).

In the special case of $\alpha \rightarrow \infty$, the game playing defined by the constructive generalization~(\ref{construct-new-strategy})
	is the same in principle as fictitious play introduced by G.W. Brown in 1951~\cite{Brown1951}.
In fictitious play, each player takes the optimal action(s) in respond to the strategies of other players.

\begin{definition}
Given a non-negative real value for the selfishness level $\alpha$, i.e., $\alpha \ge 0$. 
If the iterative computation defined by (\ref{construct-new-strategy}) reaches an equilibrium, 
	that is, there is a strategy profile $p^{*}$ satisfying
\begin{equation}
p^{*}_i (x_i) = \frac{\left(u_i (x_i, p^{*}_{-i}))\right)^{\alpha}}{ \sum_{x_i \in S_i} \left(u_i (x_i, p^{*}_{-i})\right)^{\alpha}}  \quad \mbox{for $i=1,2,\ldots, n$} \ , 
\label{general-equilibrium}
\end{equation}
then the strategy profile $p^{*}$ is called a generalized equilibrium.
\end{definition}

In parallel with Nash's 1950 PNAS paper, 
	the proof of the existence of a generalized equilibrium given any selfishness level is provided below.
Furthermore, it will be shown that when the selfishness level is sufficiently high,
	a generalized equilibrium falls back to a Nash equilibrium.

%%%%%%%%%%%%%%%%%%%%%%%%%%%%%%%%%%%%%%%%%%%%%%%%%%%%%%%%%%%%%%%%%%%%%%%%%%%%%%%%%%%%
\begin{theorem}
A generalized equilibrium $p^{*}$ defined by (\ref{general-equilibrium}) exists for any $n$-player normal form game
	with any selfishness level $\alpha$ of a non-negative value $(\alpha \ge 0)$.
It is still true even if each player $i$ in the game has his own selfishness level $\alpha_i$, possibly different from the rest.
\end{theorem}
%%%%%%%%%%%%%%%%%%%%%%%%%%%%%%%%%%%%%%%%%%%%%%%%%%%%%%%%%%%%%%%%%%%%%%%%%%%%%%%%%%%%
\begin{proof}
The set of iterative equations~(\ref{construct-new-strategy})
	defines a mapping from the strategy profile set to itself.
Because the set is compact and the mapping is continuous, 
	so a fixed point exists based on Brouwer fixed point theorem.
\end{proof}

The second part of this theorem tells us that, for any $n$-player normal form game,
	even if the selfishness level is different from player to player,
	a generalized equilibrium still exists for the game.

It is important to note that (\ref{general-equilibrium}) defines a system of polynomials if $\alpha$ is an integer.
If it is also an even number, then any real value solution to this system,
	which must also be a positive solution, 
	is also a generalized equilibrium for the game playing.
Also, the game playing defined by (\ref{construct-new-strategy})
	can be treated as an iterative, direct method to find an equilibrium of the game playing.
It defines a protocol for the players in a game
	to interact with each other so that an equilibrium can be reached with any selfishness level.
Following this protocol, each player only needs to know his own utility function 
	and the strategies of other players at the current time to compute his strategy for the next time.
The strategies of other players can be obtained through either statistical learning or message passing among the players.

%%%%%%%%%%%%%%%%%%%%%%%%%%%%%%%%%%%%%%%%%%%%%%%%%%%%%%%%%%%%%%%%%%%%%%%%%%%%%%%%%%%%
\begin{theorem}
When the selfishness level $\alpha$ is sufficiently large, i.e., 
	$\alpha \rightarrow \infty$, 
	any generalized equilibrium defined by (\ref{general-equilibrium})
	can be arbitrarily close to a Nash equilibrium and {\it vice versa}.
\end{theorem}
%%%%%%%%%%%%%%%%%%%%%%%%%%%%%%%%%%%%%%%%%%%%%%%%%%%%%%%%%%%%%%%%%%%%%%%%%%%%%%%%%%%%

The proof is given in the subsection~\ref{proof-theorem2} in the Appendix.

As a consequence, any real value solution to the system of polynomials defined by (\ref{general-equilibrium})
	with a large even number for $\alpha$ can be served as a good approximation to a Nash equilibrium.
Alternatively, the game playing defined by (\ref{construct-new-strategy}) with a sufficiently large $\alpha$
	can be applied directly to reach an equilibrium which can also be served 
	as a good approximation of a Nash equilibrium.

When $\alpha \rightarrow \infty$,
	from (\ref{construct-new-strategy}) we can see that 
	$p_i(x_i, t+1)$ is no longer a continuous function of $u_i(x_i, p_{-i}(t))$.
In this case, any mixed strategy can be extremely unstable for the slightest change in $u_i(x_i, p_{-i}(t))$
	caused by the inaccuracy at representing the utility functions,
	the variation of the utility functions,
	any perturbation and communication error among the players.
For example, a small variation in the utility function could lead to a dramatic shift of the equilibrium
	from one point in the strategy profile space to another one.
It is hard for an algorithmic method to converge to an unstable equilibrium purely based on iterations.
	
Even if a game of completely selfish players can reach an equilibrium,
	 it may have an enormous number of equilibria, 
	 possibly growing exponentially with the number of the players of the game.
The players could get stuck into one Nash equilibrium or another,
	depending on the initial conditions and sensitive to perturbations.
How to reach an equilibrium which gives relatively good overall payoff for the game becomes a challenging problem
(the overall payoff for a game is defined as the summation of the players' payoffs, i.e., $\sum_i u_i(x, t)$).

As a summary, we can say that complete selfishness of the players in a game 
	may lead to the difficulty for the players to reach an equilibrium.
Even if an equilibrium is found, it could also be unstable, sensitive to perturbations, 
	sensitive to inaccuracy or variations in utility functions, 
	and vulnerable to communication errors.
Furthermore, the overall payoff of the game may be ignored due to the fact that each player only tries to maximize his own payoff.
It is desirable to improve the overall payoff for a society 
	because it stands for improved individual payoff on average.
Also, everyone in the society could benefit from the improved overall payoff if 
	some social welfare system is implemented to redistribute the social wealth.	
Can we improve the overall payoff and the stability of a game playing 
	by simply reducing the selfishness level of the players in the game?

Both experimental result in the following section 
	and a theoretical investigation in Appendix (subsection~\ref{theoretical-investigation})
	will affirm the above question.
The theoretical investigation shows that	
	the game playing defined by the constructive generalization~(\ref{construct-new-strategy})
	is a variation of a global optimization algorithm~\cite{HuangBookCCO} defined by a multi-agent system.
When the value of the parameter $\alpha$ is reduced below a certain threshold,
	the global optimization algorithm has one and only one equilibrium and converges to it with an exponential rate.
If the equilibrium is also a consensus one among all the agents, then it must be the global optimum, guaranteed by theory.
Above the threshold, the number of equilibria of the global optimization algorithm
	may grow with the value of the parameter $\alpha$.
That is, the algorithm becomes less stable with the increase of $\alpha$,
	but the chance of reaching a consensus increases, however.

The theory suggests that
	reducing the selfishness level from the extreme of complete selfishness can stabilize the game playing
	and possibly improve the overall payoff. 
The experiments in the next section verifies that the overall payoff for many games
	is best in a statistical sense at a certain level of selfishness,
	neither at the complete selfishness one nor at the complete selfishlessness one.
Applying this to social situations, it suggests that a society should let some level of selfishness remain in its individuals. 
Otherwise, nobody has any motivation to pursue better payoffs.
Also, it is not recommended to take the other extreme where everyone is completely selfish.
However, how to find the best selfish level for any game remains as an open question.

%------------------------------------------------------
\section{Experimental Results}
%------------------------------------------------------

The prisoner's dilemma constitutes a basic problem in game theory.
It is a typical non-zero-sum game in which two players can either ``cooperate'' or ``defect'' the other player.
In this game, the only concern of each individual player (``prisoner'') is to maximize his/her own payoff.
Regardless of what the opponent chooses, 
	each prisoner always receives a higher payoff by defecting; 
		i.e., defecting is the strictly dominant strategy.
Therefore, the only possible Nash equilibrium for the game  is for all prisoners to defect. 

An example payoff matrix of the prisoner's dilemma is given as follows:

\begin{center}
\begin{tabular}{|c|c|c|}
 \multicolumn{1}{c}{}& \multicolumn{1}{c}{Cooperate} &  \multicolumn{1}{c}{Defect}\\
\cline{2-3}
\multicolumn{1}{r|}{Cooperate} &  ~3,3~ & ~1,4~\\
\cline{2-3}
\multicolumn{1}{r|}{Defect} & ~4,1~ & {\bf 2,2} \\ 
\cline{2-3}
\end{tabular}
\end{center}

\begin{figure}
\centering
\center{\epsfxsize 9.8cm \epsffile{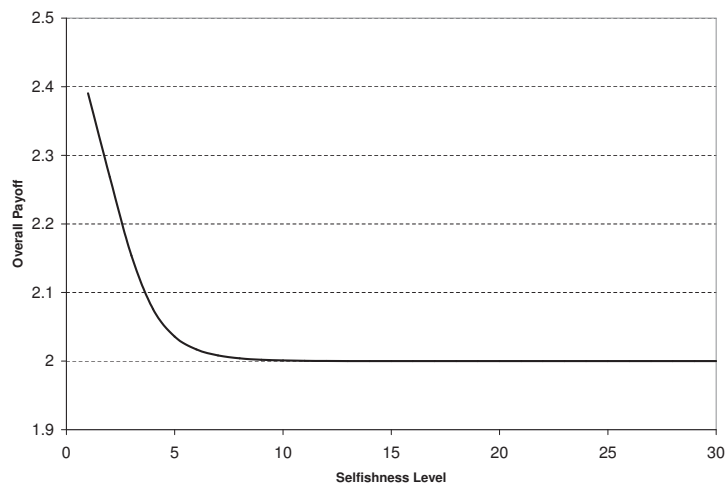}}
\caption{Payoffs of prisoner's dilemma under different selfishness levels.}
\label{fig-1}
\end{figure}

At the Nash equilibrium (the element in the matrix with a bold font), 
	the payoffs of the two players are (2,2).
It corresponds to the case when the selfishness level $\alpha = \infty$.
When the two players reduce their selfishness level together, 
	their payoffs at equilibria also increase together as shown in Figure~\ref{fig-1}.
Those equilibria are found 
	by the constructive generalization~(\ref{construct-new-strategy}) with different selfishness levels.

From Fig.~\ref{fig-1} we can see that when the two players have the same selfishness level 
	and the level is of a high value ($\alpha = 30$),
	their payoffs are close to those of the Nash equilibrium.
The moment that the both players reduce their selfishness level, 
	both get better payoffs than those of the Nash equilibrium.
When the selfishness level reduces to one ($\alpha = 1$), 
	the payoffs are close to $2.4$ for both, a $20\%$ increases over the one of the Nash equilibrium.
	
%Often times in the real world we face a situation similar to the prisoner's dilemma.
%If each of us only considers our own payoff, showing no interest at the payoffs of others, 
%	we could end up with less payoffs than the case when we together reduce our selfishness level
%	and show some interest at the payoffs of others.
	
The result seems counterintuitive because if one player could update his strategy to improve his payoff,
	he should go ahead to do it in order to receiving a better payoff.
However, in many cases, all the players in a game are inter-connected.
The gain of one player often leads to the loss of other players.
If everyone yields back a little bit of his payoff as a favor to others, 
	everyone can end up with better payoff as a returned favor from others instead.

To show the power of the constructive generalization at finding Nash equilibria, 
	a 2-player game is used with the following payoff matrix:
	
\[
\left(
\begin{array}{rrrrr}
 2,~3~ & ~-1,~4~ & ~2,~4~ & ~5,~2~ & ~1,-1 \\
 2,~2~ & ~3,~0~ & ~4,~1~ & ~-2,~4~ & ~1,~3 \\
 4,~6~ & ~7,~2~ & ~2,-2~ & ~4,~9~ & ~2,~1 \\
 9,~0~ & ~-2,~6~ & ~6,~3~ & ~7,~0~ & ~0,~5 \\
 3,~2~ & ~6,~1~ & ~2,~5~ & ~5,~3~ & ~1,~0 \\
\end{array}
\right)
\]

This game has been used in other game theory literatures as a hard game
	because it has only one mixed Nash equilibrium.
The strategy for the row player is $(0,0, \frac{2}{11}, \frac{4}{11}, \frac{5}{11})$ with the payoff $4$. 
The strategy for the column player is $(0, \frac{2}{7}, \frac{3}{7}, \frac{2}{7}, 0)$ with the payoff $3$.
This mixed Nash equilibrium is extremely unstable.
Assume that the two players play the game by taking only the best action.
Assume further that the column player couldn't represent fraction numbers.
Instead, the player uses real values to approximate them, just like the real values stored in most computers. 
Then, a very slight round-off error for the value $\frac{3}{7}$, say $0.4285714285714285714285714286$,
	could knock the row player off of his Nash equilibrium strategy to the new one $(0, 0, 0, 1, 0)$,
	which will in turn knock the column player off his Nash equilibrium strategy.
As a consequence, both of them will immediately be knocked off the Nash equilibrium 
	and get stuck into a chaotic situation.
Therefore, this game is hard for an iteration-based direct method to reach the unique mixed Nash equilibrium.

Despite of its hardness, 
	the constructive generalization~(\ref{construct-new-strategy})
	as an iteration-based direct method can find a very good approximation to the Nash equilibrium.
To improve the convergence property of the method, 
	an additional step is added after computing $p_i (x_i, t)$ defined by (\ref{construct-new-strategy})
	to smooth out its fluctuation.
It is done by keeping some memory of the previous value of $p_i (x_i, t)$, i.e.,
\[ \lambda p_i (x_i, t+1) + ( 1- \lambda) p_i (x_i, t) \rightarrow p_i (x_i, t+1) \ , \]
where $\lambda = 0.001$ was used in the experiment.

Furthermore, to increase the chance for the constructive generalization 
	to reach an equilibrium at a high selfishness level $\alpha$,
	the value of $\alpha$ is progressively raised from a small value, say $1$.
When it reaches the value  $1000$,
	the payoff for the row player is $4.0068$, a difference around $0.17\%$ to the payoff 4 of the Nash equilibrium.
His strategy (left) is very close to the Nash equilibrium one (right) as shown below:
\[ (0, 0, \frac{1.999}{11}, \frac{3.999}{11}, \frac{5.002}{11}) \approx (0,0, \frac{2}{11}, \frac{4}{11}, \frac{5}{11}) \ . \]
The payoff for the column player is $3.0001097$, a difference around $0.0037\%$ to the payoff $3$ of the Nash equilibrium.
His strategy (left) is very close to the Nash equilibrium one (right) as shown below:
\[ (0, \frac{1.997}{7}, \frac{2.981}{7},  \frac{2.023}{7}, 0) \approx (0, \frac{2}{7}, \frac{3}{7}, \frac{2}{7}, 0) \ . \].

\begin{figure}
\centering
\center{\epsfxsize 9.8cm \epsffile{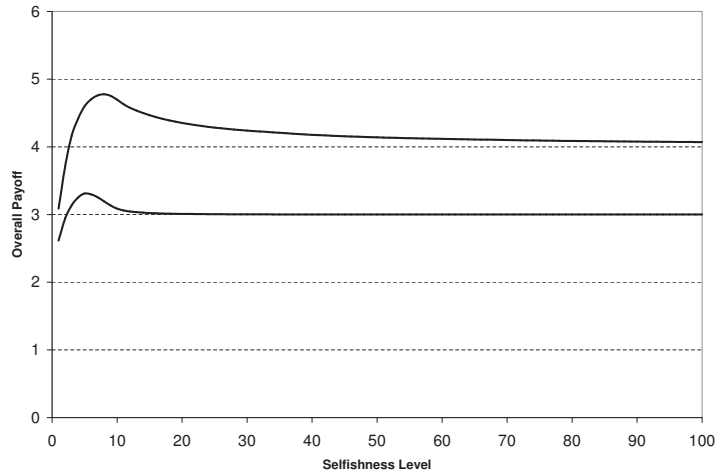}}
\caption{Payoffs of two players with 5 actions under different selfishness levels.}
\label{fig-2}
\end{figure}

Fig.~\ref{fig-2} shows the changes of the payoffs of the two players in relation to the selfishness level.
The payoff of the row player is peaked around $\alpha=8$ with the value $4.77572$,
The payoff of the column player is peaked around $\alpha=5$ with the value $3.30341$.
At $\alpha = 7$, the payoffs for both are $(4.7586,3.2527)$, a $19\%$ improvement for the row player 
	and a $8.4\%$ improvement for the column player over the payoffs $(4,3)$ of the Nash equilibrium.

To illustrate the power of the constructive generalization~(\ref{construct-new-strategy}) at stabilizing game playing,
	a 2-players game with 6 actions for each is constructed with the following payoff matrix:
\[
\left(
\begin{array}{ccccccc}
{\bf 6,~6}~ & ~1,~1~ & ~1,~1~ & ~1,~1~ & ~1,~1~ & ~1,~1 \\
 1,~1~ & ~{\bf 6.5,~6.5}~ & ~1,~1~ & ~1,~1~ & ~1,~1~ & ~1,~1 \\
 1,~1~ & ~1,~1~ & ~{\bf 7,~7}~ & ~1,~1~ & ~1,~1~ & ~1,~1 \\ 
 1,~1~ & ~1,~1~ & ~1,~1~ & ~{\bf 7.5,~7.5}~ & ~1,~1~ & ~1,~1 \\ 
 1,~1~ & ~1,~1~ & ~1,~1~ & ~1,~1~ & ~{\bf 8,~8}~ & ~1,~1 \\ 
 1,~1~ & ~1,~1~ & ~1,~1~ & ~1,~1~ & ~1,~1~ & ~{\bf 8,~8.5} \\ 
\end{array}
\right)
\]

Clearly, this game has six actions for each player and six pure Nash equilibria.
Let us label the actions for each player as $1,2,3,4,5,6$.
If any player picks the $i$th action in random,
	the other will take the same action as the best response.
As a consequence, a Nash equilibrium is thus found.

\begin{figure}
\centering
\center{\epsfxsize 10cm \epsffile{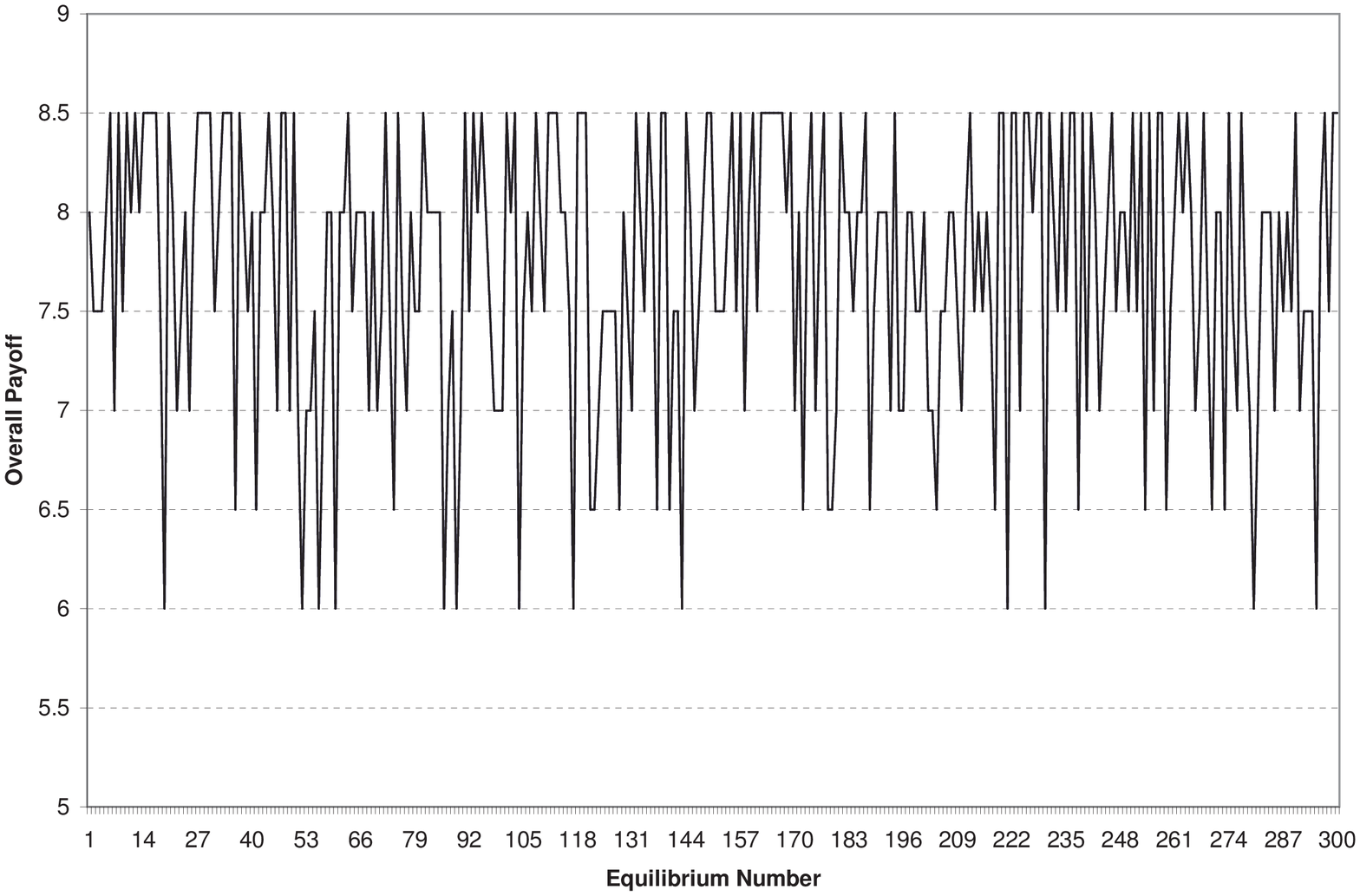}}
\caption{Overall payoffs of 300 generalized Nash equilibria with the selfishness level $\alpha= 100$. The average $\mu = 7.728$ and variance $\sigma^2=0.502$.}
\label{fig_3}
\end{figure}

\begin{figure}
\centering
\center{\epsfxsize 10cm \epsffile{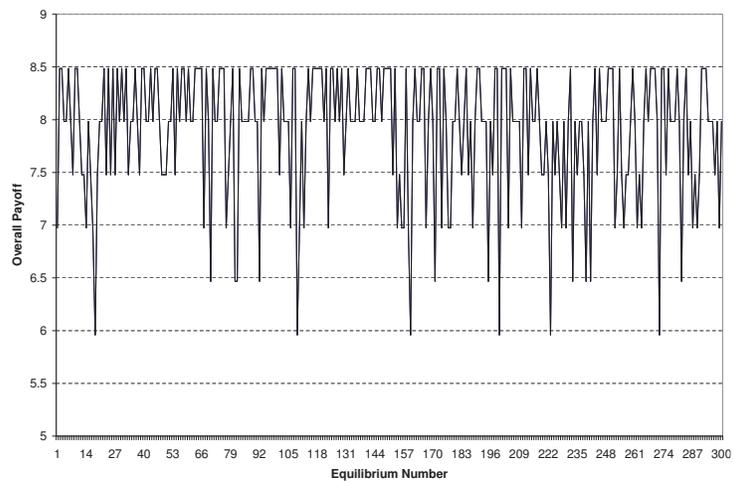}}
\caption{Overall payoffs of 300 generalized Nash equilibria with the selfishness level $\alpha= 4$. The average $\mu = 7.910$ and variance $\sigma^2=0.379$.}
\label{fig_4}
\end{figure}

\begin{figure}
\centering
\center{\epsfxsize 10cm \epsffile{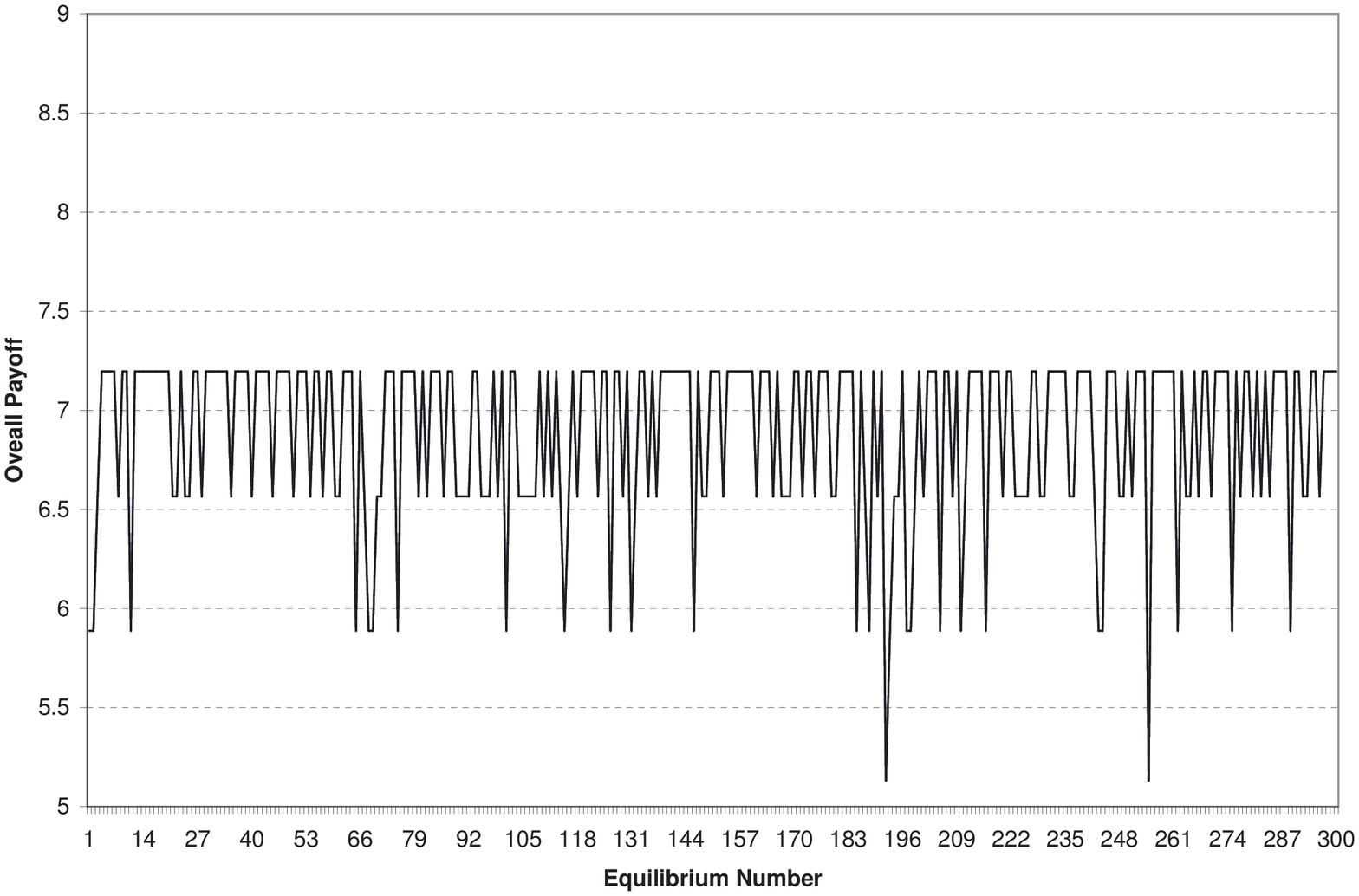}}
\caption{Overall payoffs of 300 generalized Nash equilibria with the selfishness level $\alpha= 2$. The average $\mu = 6.888$ and variance $\sigma^2=0.193$.}
\label{fig_5}
\end{figure}

With this best-response playing, the average payoff for each player is $7.725$,
	the variance of the payoff is $35/48 \approx 0.729$.
Three hundred generalized equilibria are found 
	using the constructive generalization~(\ref{construct-new-strategy}) 
	with the selfishness levels $\alpha=100,4,2,1$, respectively.
The results for $\alpha=100,4,2$ are shown in Fig.~\ref{fig_3}, Fig.~\ref{fig_4}, and Fig.~\ref{fig_5} respectively.
From the figures we can see that the stability of the game playing defined by the constructive generalization 
	improves progressively as the selfishness level $\alpha$ decreases.
Here, the stability is reversely proportional to the variance of the payoff.
When $\alpha = 1$, the game playing always converges to a unique equilibrium with the payoff=$2.10373$ after three hundred runs.
That is, the game playing tends to have only one equilibrium when the selfishness level drops below a certain threshold.
Also we can see from the three figures that the average payoff of each player is of the highest value when $\alpha=4$.

From this example, we can see that the constructive generalization yields the best payoffs for the players in a game
	at a certain selfishness level.
The stability of the game playing continuously improves as the selfishness level reduces.
That is, reducing the selfishness level can always improve the stability of the game playing.
However, the players in a game can only get the highest payoffs at a statistical sense at a certain selfishness level
(In the subsection~\ref{theoretical-investigation} in the Appendix, a theoretical investigation is given to offer some explanation).

In the following set of experiments, computer-generated societies with a population ranging from hundreds to a thousand
	are used to demonstrate the improvement of payoffs and stability by reducing the selfishness level.
In each society, each individual has a number of neighbors
	and his payoff function is defined by the summation of the pairwise joint actions of himself and his neighbors as follows
\begin{equation}
u_i (x) = \sum_{j \in {\cal N}(i)} f_{ij} (x_i, x_j) \ , 
\label{binary_payoff_function}
\end{equation}
where ${\cal N}(i)$ is the set of the individual $i$'s neighbors.
The overall payoff of the society is defined as
\[ \sum_i u_i(x) = \sum_i \sum_{j \in {\cal N}(i)} f_{ij} (x_i, x_j) \ . \]

Each function value $f_{ij}(x_i, x_j)$ is uniformly sampled from the interval $[0, 1]$.
The neighbors of each individual are randomly picked from the entire population.

\begin{figure}
\centering
\center{\epsfxsize 10cm \epsffile{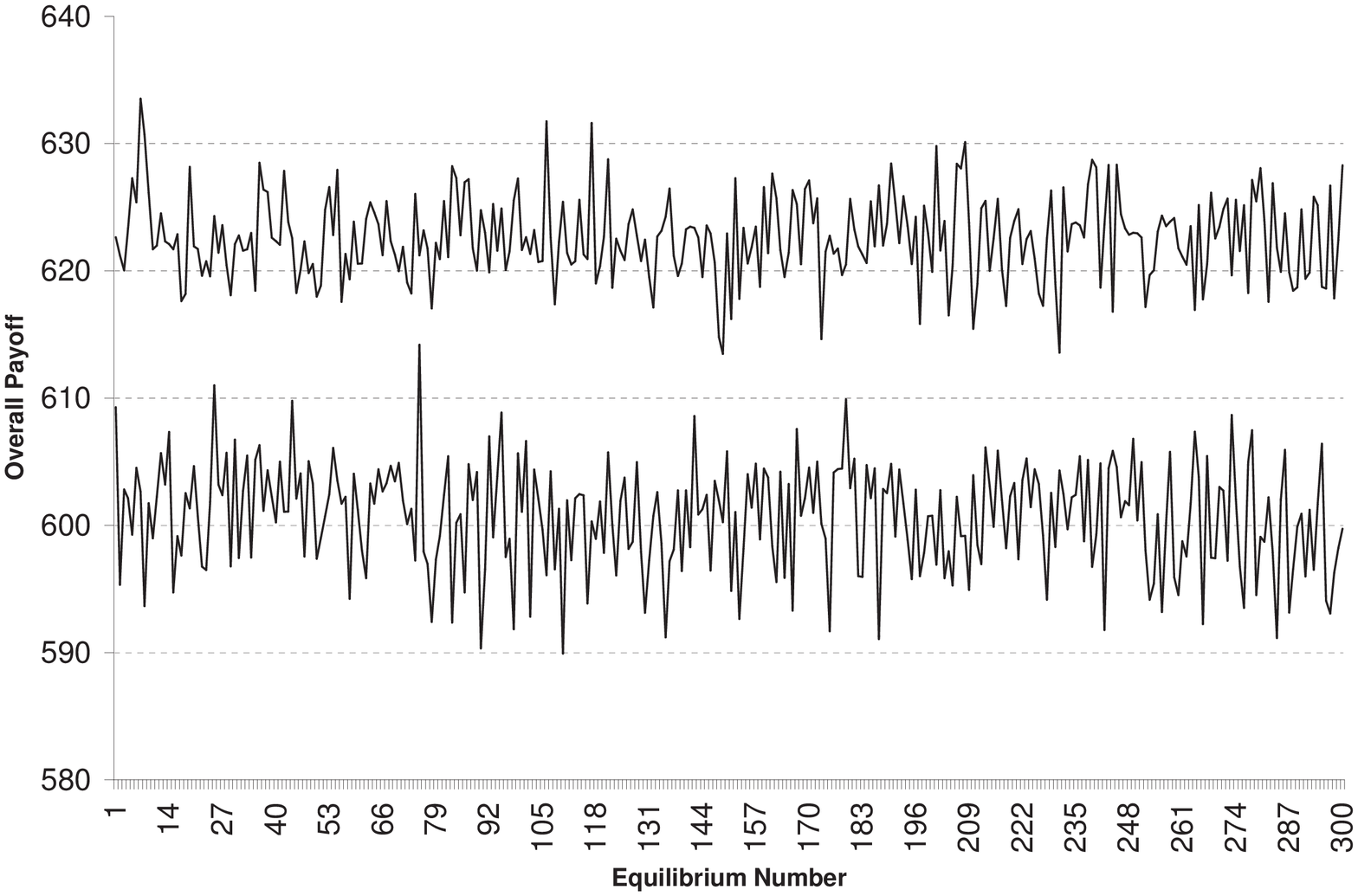}}
\caption{Overall payoffs of 300 Nash equilibria (bottom) versus 300 generalized ones (top, the selfishness level $\alpha=20$) 
	for a society of $121$ individuals.
For the former, the average $\mu = 600.67$ and variance $\sigma^2=17.5$.
For the latter, the average $\mu = 622.60$ and variance $\sigma^2=10.7$.}
\label{fig_6}
\end{figure}

In the first experiment,
	an instance of a society of $121$ individuals is generated where each one has $50$ actions and $6$ neighbors on average.
$300$ Nash equilbria are discovered by fictitious play and $300$ generalized ones 
	 are discovered by the constructive generalization with the selfishness level $\alpha=20$.
Fig.~\ref{fig_6} shows the overall payoffs of the first 300 ones versus the second 300 ones.
From the figure we can see that, reducing the selfishness level can lead to remarkable improvement both in payoffs and stability.

\begin{figure}
\centering
\center{\epsfxsize 10cm \epsffile{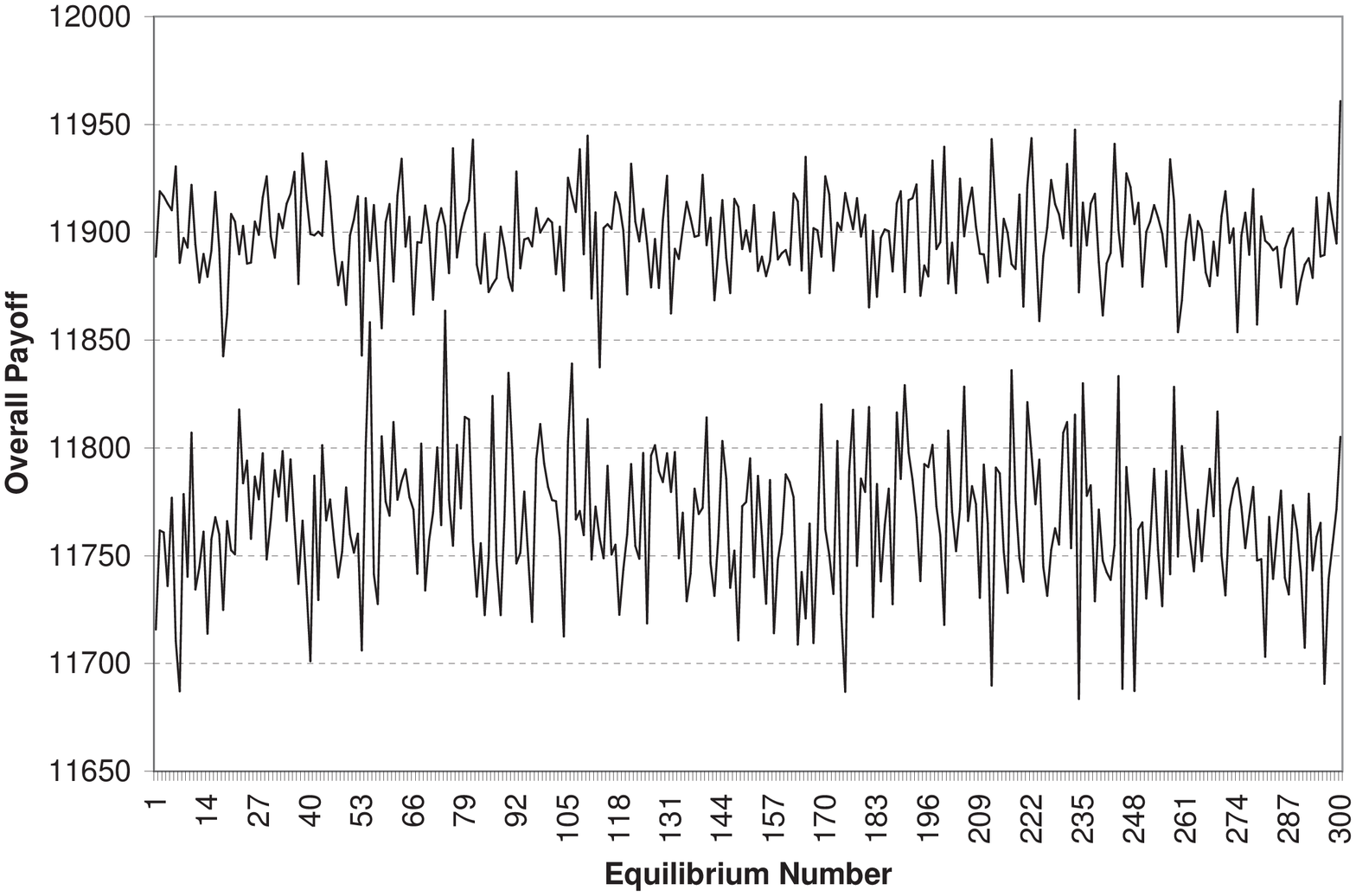}}
\caption{Overall payoffs of 300 Nash equilibria (bottom) versus 300 generalized ones (top, the selfishness level $\alpha=20$) 
	for a society of $601$ individuals.
For the former, the average $\mu = 11766$ and variance $\sigma^2=1009$.
For the latter, the average $\mu = 11899$ and variance $\sigma^2=392$.
}
\label{fig_7}
\end{figure}

In the second experiment, the population is increased to $601$, the number of actions per person is reduced to $20$,
	and the size of neighbors on average is increased to $30$.
Figure~\ref{fig_7} shows the overall payoffs of 300 Nash equilibria versus the 300 generalized ones
	with the selfishness level = $20$.
From the figure we can see that, reducing the selfishness level can lead to remarkable improvement both in payoffs and stability
	with a larger population.

\begin{figure}
\centering
\center{\epsfxsize 10cm \epsffile{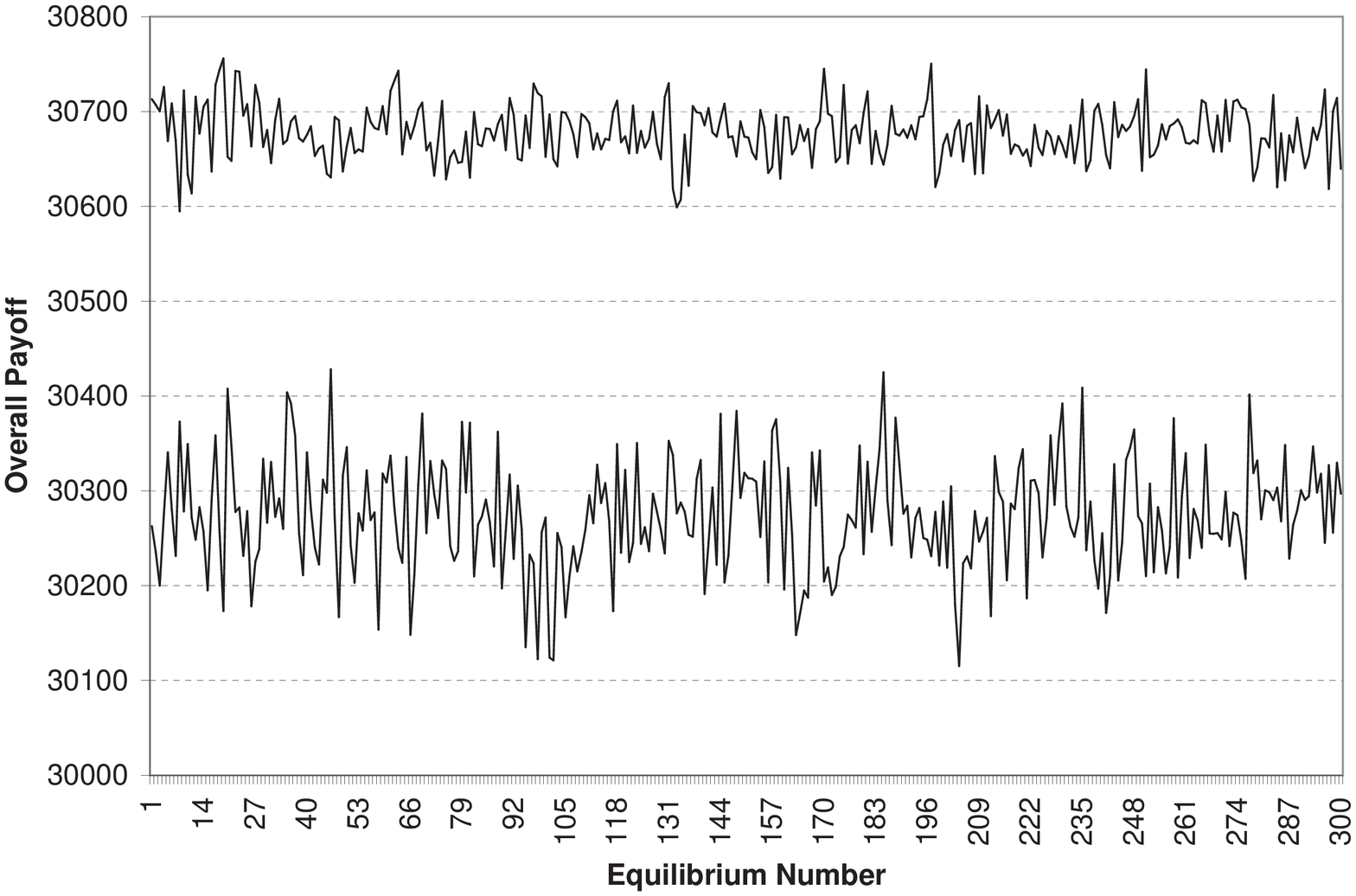}}
\caption{Overall payoffs of 300 Nash equilibria versus 300 (bottom) generalized ones (top, the selfishness level $\alpha=30$) 
	for a society of $1001$ individuals.
For the former, the average $\mu = 30274$ and variance $\sigma^2=3335$.
For the latter, the average $\mu = 30677$ and variance $\sigma^2=818$.
}
\label{fig-8}
\end{figure}

In the third experiment, the population is increased further to $1001$, the number of actions per person is reduced to $10$,
	and the size of neighbors on average is increased to $50$.
From Figure~\ref{fig-8} we can make the same conclusions as above with an even larger population.
	
The last three experiments with societies of different population sizes
	are extended with more selfishness levels.
The average overall payoff and the fluctuation of the overall payoff of a society with different selfishness levels $\alpha$ 
	are shown in the following table.
The fluctuation is indicated 
	by the variance of the overall payoff given a selfishness level.
The less fluctuation a society has, the more stable the society is.
	
\begin{center}
\begin{tabular}{|ccc|cc|cc|}
\hline
Population Size & \multicolumn{2}{c}{121} & \multicolumn{2}{c}{601} & \multicolumn{2}{c|}{1001} \\
\hline
\multicolumn{1}{|c|}{Selfishness Level}&~Payoff~&~Fluctuation~&~Payoff~&~Fluctuation~&~Payoff~&~Fluctuation~ \\
\hline
\multicolumn{1}{|c|}{$\infty$} & 601 & 17.5 & 11766 & 1009 & 30274 & 3335 \\
\multicolumn{1}{|c|}{100} & 611 & 16.9 & 11883 & 591 & 30487 & 1371 \\
\multicolumn{1}{|c|}{80} & 611 & 15.7 & 11902 & 516 & 30575 & 1296 \\
\multicolumn{1}{|c|}{60} & 613 & 13.2 & 11939 & 471 & 30523 & 1151\\
\multicolumn{1}{|c|}{50} & 615 & 13.0 & 11956 & 461 & 30612 & 1069\\
\multicolumn{1}{|c|}{40} & 617 & 10.1 & 11983 & 455 & 30649 & 1029\\
\multicolumn{1}{|c|}{30} & 619 & 10.1 & {\bf 12014} & 411 & {\bf 30677} & 818\\
\multicolumn{1}{|c|}{20} & {\bf 623} & 8.68 & 11899 & 392 & 30360 & 681\\
\multicolumn{1}{|c|}{10} & 458 & 0 & 10276 & 0& 27988 & 0\\
\hline
\end{tabular}
\end{center}

From the above table, we can see that 
	the overall payoffs of the three societies improve progressively with the reduction of the selfishness level $\alpha$
	started from $\alpha = \infty$ (complete selfishness).
Each society yields the highest overall payoff at a some selfishness level and degrades progressively 
	with further reduction of the selfishness level.
The stability of each society continuously improves as the selfishness level reduces.
That is, reducing the selfishness level can always improve the stability of a society.
This experiment shows us that 
	a less selfish society can be better in overall payoff and stability
	than a completely selfish society. 

A less selfish society can also be more efficient 
	than a completely selfish society.
The efficiency of a society can be measured 
	by the capability at finding a good equilibrium in terms of the overall payoff.
To compare the efficiency, 
	the same society of a population of $121$ described before is used in the experiment.
When the individuals in the society are less selfish ($\alpha = 20$),
	the average overall payoff of the 300 equilibria found by the society is $622.60$ (see also Fig.~\ref{fig_6}).
When all the individuals become completely selfish,
	after exploring one million of equilibria by the society, 
	the best overall payoff is of a value $621.5$, less than the former one $622.60$.
This result says that the average payoff of the less selfish society in a generalized equilibrium 
	is better than the best payoff out of those of one million Nash equilbria explored by the completely selfish society.
The less selfish society spent seconds on average to find an equilibria
	while the completely selfish society took almost a whole day to find the one million equilibria using a laptop with a AMD Turion\texttrademark X2 Dual-Core Mobile Processor and 3GB RAM.
The less selfish society is several orders of magnitude more efficient than the completely selfish society.

\begin{figure}
\centering
\center{\epsfxsize 10cm \epsffile{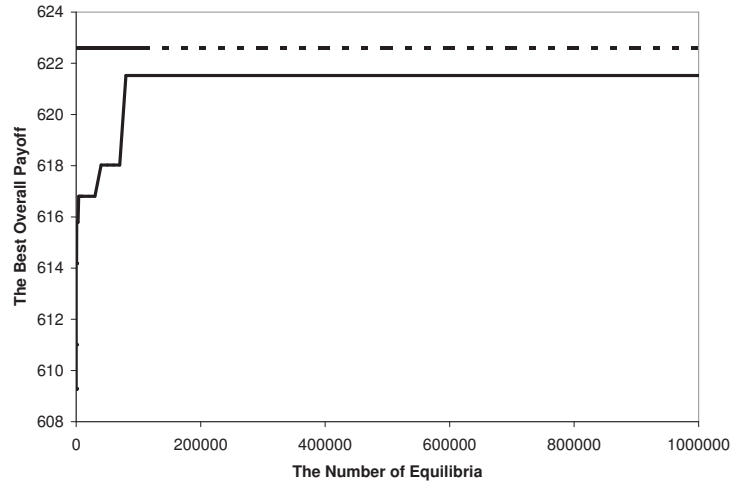}}
\caption{After exploring one million equilibria by a society of $121$ completely selfish individuals, 
		the best one in terms of overall payoff still couldn't match the average one (dotted line) found by the same society when all the individuals are less selfish.}
\label{fig-9}
\end{figure}

Fig.~\ref{fig-9} shows the improvement of the best overall payoff
	with the increase of the number of equilibria discovered by the completely selfish society mentioned above.
	
%------------------------------------------------------
\section{Conclusions}
%------------------------------------------------------

John Nash in his Nobel price-winning work defined an equilibrium and proved its existence for $n$ players games
            where all players are completely selfish.
However, it is important from both a theoretical and a practical point of view 
	to understand game playing where players are less selfish.
The key contribution of this paper is a generalization of Nash equilibrium to cover
            the entire spectrum of selfishness ranging from complete selfishness to complete selfishlessness.
%The key contribution of this paper is a generalization of Nash equilibrium 
%	along the line of the selfishness level described by a mathematical model.
It also gives the proof of the existence of an equilibrium for a game of $n$-players with any selfishness level.
The definition of Nash equilibrium is a special case of this generalization 
	where all players are completely selfish.
The generalization is constructive since it offers a protocol for players in a game to reach an equilibrium.
Most importantly, this paper presents experimental results and theoretical investigation 
	to show that the players in a game can reduce their selfishness level together
	to reach a new equilibrium where they can have better payoffs
	and the game playing is more stable at the same time.

\nocite{GameTheoryLuce}

\nocite{Holt04}
%\nocite{DaskalakisGoldbergPapadimitriou05}
%\nocite{Chen&Deng06}
\nocite{Lemke&Howson64,Laan87,Govindan03}
\nocite{HuangBookCCO}
%\bibliography{../bib/AIsfs}

%----------------------------------------------------------------------------------------------------
\section{Appendix}
%----------------------------------------------------------------------------------------------------

%----------------------------------------------------------------------------------------------------
\subsection{Proof for Theorem~2}
\label{proof-theorem2}
%----------------------------------------------------------------------------------------------------

%------------------------------------------
\subsubsection {Definitions and Notations}
%------------------------------------------

At time instance $t$, 
	let $u_i (x_i, p_{-i}(t))$ be the payoff of player $i$ 
	by taking action $x_i$ in response to other players' strategies $p_{-i}(t)$. 
It is a function of $x_i$ and $t$, called the action payoff function, denoted as $\Psi_i(x_i, t)$. 
Obviously, we have
\begin{equation}
\Psi_i (x_i, t) = u_i (x_i, p_{-i}(t))=  \sum_{\sim x_i} \left( u_i(x) \prod_{j \not= i} p_j (x_j, t)\right),\quad \mbox{for any $i$} \ . 
\label{action-payoff-function}
\end{equation}

Using the notation,
	the constructive generalization~(\ref{construct-new-strategy}) can be rewritten as 
\begin{equation}
p_i (x_i, t+1) = \frac{\left(\Psi_i (x_i,t)\right)^{\alpha}}{ \sum_{x_i \in S_i} \left(\Psi_i (x_i,t)\right)^{\alpha}},  \quad \mbox{for $i=1,2,\ldots, n$} \ . 
\label{construct-new-strategy-a}
\end{equation}
That is, $p_i (x_i, t+1)$ equals to the normalized $(\Psi_i (x_i,t))^{\alpha}$.
To show the relationship, $p_i(x_i, t)$ can be expressed as $\left({\bar \Psi}_i(x_i,t)\right)^{\alpha}$ 
	with the bar standing for the normalization.
That is,
\begin{equation}
p_i (x_i, t+1) = \left({\bar \Psi}_i (x_i,t)\right)^{\alpha},  \quad \mbox{for $i=1,2,\ldots, n$} \ . 
\label{construct-new-strategy-b}
\end{equation}
Substituting (\ref{construct-new-strategy-b}) into (\ref{action-payoff-function}),
	we have an iterative update function for $\Psi_i(x_i, t)$ as follows
\begin{equation}
\Psi_i (x_i, t+1) = \sum_{\sim x_i} \left( u_i(x) \prod_{j \not= i} \left({\bar \Psi}_i (x_i,t)\right)^{\alpha} \right),\quad \mbox{for $i=1,2,\ldots,n$} \ . 
\label{cooperative_optimization_general3a}
\end{equation}

If a strategy profile $p^{*}$ is a generalized equilibrium satisfying (\ref{general-equilibrium}),
 then there is a corresponding set of action payoff functions 
	$\{ \Psi^{*}_1(x_1), \Psi^{*}_2(x_2), \ldots, \Psi^{*}_n(x_n) \}$ defined by (\ref{action-payoff-function}), 
	or simply $\Psi^{*}$, such that
	(\ref{cooperative_optimization_general3a}) is satisfied. 
That is,
\begin{equation}
\Psi^{*}_i (x_i) = \sum_{\sim x_i} \left( u_i(x) \prod_{j \not= i} \left({\bar \Psi}^{*}_j (x_j)\right)^{\alpha} \right),
	\quad  \mbox{for $i=1,2,\ldots, n$} \ . 
\label{generalized-equilibrium}
\end{equation}
Both a strategy profile $p^{*}$ satisfying (\ref{general-equilibrium}) and
	an action payoff function set $\Psi^{*}$ satisfying (\ref{generalized-equilibrium}) can be used to represent a generalized equilibrium.
Based on (\ref{action-payoff-function}), we have
\[ \Psi^{*}_i (x_i) = \sum_{\sim x_i} \left( u_i(x) \prod_{j \not= i} p^{*}_j (x_j)\right) ,\quad \mbox{for $i=1,2,\ldots,n$}\ . \]
Based on (\ref{construct-new-strategy-b}), we have
\[ p^{*}_i (x_i) = \left({\bar \Psi}^{*}_i (x_i)\right)^{\alpha} ,\quad \mbox{for $i=1,2,\ldots,n$}\ . \]

%-----------------------------
\subsubsection{The Proof}
%-----------------------------

The best action of player $i$ at time $t$ is defined as the one with the highest payoff,
	i.e., the $x_i$ that maximizes the action payoff function $\Psi_i(x_i, t)$.
Assume that the total number of actions of player $i$ is $m_i$.
Assume further that $\alpha \ge 1$.
At a generalized equilibrium with a strategy profile $p^{*}$ and its corresponding action payoff function set as $\Psi^{*}$, 
	based on (\ref{generalized-equilibrium}), 
	we can find out the difference between the best payoff $\max_{x_i} \Psi^{*}_i(x_i)$ 
	and the expected payoff $\sum_{x_i} \Psi^{*}_i(x_i)  p^{*}_i (x_i)$.
It is straightforward to verify that the difference should satisfy the following inequality:

\[ 0 \le \max_{x_i} \Psi^{*}_i(x_i) - \sum_{x_i} \Psi^{*}_i(x_i)  p^{*}_i (x_i)< \left(\frac{m_i - 1}{e} \max_{x_i} \Psi^{*}_i(x_i)\right) \alpha^{-1}\ . \]
Obviously, the difference can be arbitrarily small when the parameter $\alpha$ is sufficiently large.
That is, the difference is reduced to zero when $\alpha \rightarrow \infty$,
\begin{equation}
\lim_{\alpha \rightarrow \infty} \left( \max_{x_i} \Psi^{*}_i(x_i) - \sum_{x_i} \Psi^{*}_i(x_i) p^{*}_i (x_i) \right) = 0, \quad \mbox{for any $i$} \ . 
\label{cooperation_equilibrium}
\end{equation}

Given a strategy profile $p^{*}$, 
	it is a Nash equilibrium if and only if, 
	given any player, its best payoff is equal to its expected payoff $\sum_{x_i} \Psi^{*}_i(x_i) p^{*}_i (x_i)$.
That is, for any $i$,
\begin{equation}
\max_{x_i} \Psi^{*}_i(x_i) - \sum_{x_i} \Psi^{*}_i(x_i) p^{*}_i (x_i) = 0 \ . 
\label{Nash_equilibrium}
\end{equation}

Compare the statement~(\ref{cooperation_equilibrium}) with the statement~(\ref{Nash_equilibrium}), 
	we can conclude that
	any generalized equilibrium~(\ref{generalized-equilibrium}) 
	can be arbitrarily close to a Nash equilibrium if the parameter $\alpha$ is sufficiently large.

The other way around is also true. 
That is, for any Nash equilibrium, 
	there exists a generalized equilibrium defined as (\ref{generalized-equilibrium}) 
	which is arbitrarily close to the Nash equilibrium if the parameter $\alpha$ is sufficiently large.
To prove this statement, recall that the action payoff function $\Psi_i(x_i)$ computed by (\ref{cooperative_optimization_general3a})
	is the payoff of player $i$ taking the action $x_i$ while other players taking the strategies $p_j$ ($j \not = i$).
Assume that a strategy profile $p^{*}$ is a Nash equilibrium.
Then the payoff $\Psi^{*}_i(x_i)$ at the Nash equilibrium should satisfy the following condition,
\begin{eqnarray*}
\max_{x_i} \Psi^{*}_i(x_i) = \Psi^{*}_i(x_i), &\quad & \mbox{if $p^{*}_i(x_i) > 0$} \ ;  \\
\max_{x_i} \Psi^{*}_i(x_i) \le \Psi^{*}_i(x_i), &\quad & \mbox{if $p^{*}_i(x_i) = 0$ \ . }
\end{eqnarray*}

Let $\epsilon$ is a positive infinidesmal.
Note that for any probability $p_i$, if $0 < p_i \le 1$, then 
\[ \lim_{\epsilon \rightarrow 0^{+}} (1 + \epsilon \ln p_i)^{1/\epsilon} = p_i \ . \]
Otherwise, if $p_i = 0$,
then 
\[ \lim_{\epsilon \rightarrow 0^{+}} (1 + \epsilon \ln \epsilon)^{1/\epsilon} = p_i (=0) \ . \]

Given each player $i$, $i=1,2,\ldots, n$, define its action payoff function $\Psi^{'}_i(x_i)$ as
\[ \Psi^{'}_i(x_i) = \left\{ \begin{array} {l}
                               (1 + \epsilon \ln p^{*}_i(x_i)) \max_{x_i} \Psi^{*}_i(x_i),~~\mbox{if $p^{*}_i(x_i) > 0$} \ ;  \\
                               (1 + \epsilon \ln \epsilon) \max_{x_i} \Psi^{*}_i(x_i), ~~\mbox{if $p^{*}_i(x_i) = 0$ and $\max_{x_i} \Psi^{*}_i(x_i) = \Psi^{*}_i(x_i)$} \ ;  \\
                               \Psi^{*}_i(x_i), ~~\mbox{if $\Psi^{*}_i(x_i) < \max_{x_i} \Psi^{*}_i(x_i) $} \ .  
                               \end{array}
                       \right. \]       
Obviously,
\[ \lim_{\epsilon \rightarrow 0^{+}} \Psi^{'}_i(x_i) = \Psi^{*}_i(x_i) \ . \]

Let $\alpha = 1 / \epsilon$, from (\ref{construct-new-strategy-a}) 
	used for computing the strategy $p_i(x_i, t)$, we have
\[ \lim_{\epsilon \rightarrow 0^{+}} \frac{\left(\Psi^{'}_i(x_i)\right)^{ 1 / \epsilon}}{\sum_{x_i} \left(\Psi^{'}_i(x_i)\right)^{ 1 / \epsilon}} = p^{*}_i(x_i), \quad \mbox{for $i=1,2,\ldots, n$} \ . \]
Hence, the set of action payoff functions $\{ \Psi^{'}_1(x_1),\Psi^{'}_2(x_2), \ldots, \Psi^{'}_n(x_n) \}$  
	is a generalized equilibrium satisfying (\ref{generalized-equilibrium})
	when the parameter $\alpha$ is sufficiently large.
Its corresponding strategy profile
	$\{ p^{*}_1(x_1),p^{*}_2(x_2), \ldots, p^{*}_n(x_n) \}$ 
	is the strategy profile $p^{*}$ of the Nash equilibrium in the assumption.
In other words,
	for any Nash equilibrium with a strategy profile $p^{*}$, 
	there always exists a generalized equilibrium satisfying (\ref{general-equilibrium})
	which is arbitrarily close to the Nash equilibrium when the selfishness level $\alpha$ is sufficiently large.

%----------------------------------------------------------------------------------------------------
\subsection{Theoretical Investigation}
\label{theoretical-investigation}
%----------------------------------------------------------------------------------------------------

%----------------------------------------------------------------------------------------------------
\subsubsection{From Cooperative Optimization to the Constructive Generalization}
%----------------------------------------------------------------------------------------------------

The constructive generalization can be derived from a recently discovered general global optimization method, 
	called cooperative optimization~\cite{HuangBookCCO}.
Cooperation is an ubiquitous phenomenon in nature.
The cooperative optimization theory is a mathematical theory 
	for understanding cooperative behaviors and translating it into optimization algorithms.
The major theoretical results can be found in \cite{HuangBookCCO}.

Let $E(x_1, x_2, \ldots, x_n)$, or simply $E(x)$, be a multivariate objective function of $n$ variables. 
Assume that $E(x)$ can be decomposed into $n$ sub-objective functions $E_i(x)$, 
	one for each variable, such that those sub-objective functions satisfying
\[ E_1(x) + E_2(x) + \ldots + E_n(x) = E(x) \ . \]

In terms of a multi-agent system, 
	let us assign $E_i(x)$ as the objective function for agent $i$, for $i=1,2,\ldots, n$.
There are $n$ agents in the system in total.
The objective of each agent $i$ is to maximize $E_i(x)$.
The objective of the system is to maximize $E(x)$, called the global objective function.

There is a simple form of cooperative optimization
	where each agent $i$ is associated with a function $\Psi_i(x_i, t)$ 
	defined on the variable $x_i$ and time $t$.
The function is called the assignment function for the agent.
Each agent updates its assignment function iteratively as follows:
\begin{equation}
\Psi_i (x_i, t) = \sum_{\sim x_i} \left( e^{E_i(x)/\hbar} \prod_{j \not= i} p_j(x_j, t-1) \right),\quad \mbox{for $i=1,2,\ldots,n$}, 
\label{cooperative_optimization_general3}
\end{equation}
where $\sum_{\sim x_i}$ stands for the summation over all variables except $x_i$ and $\hbar$ is a constant of a small positive value.
$p_i(x_i, t)$ is defined as 
\begin{equation} 
p_i(x_i, t) = \frac{\left(\Psi_i(x_i,t)\right)^{\alpha}}{\sum_{x_i} \left(\Psi_i(x_i,t)\right)^{\alpha}} \ , 
\label{compute_assignment_probabilty}
\end{equation}
where $\alpha$ is a parameter of a non-negative real value.

By the definition, $p_i(x_i, t)$ is a probability-like function satisfying 
\[ \sum_{x_i} p_i (x_i, t) = 1 \ . \]
It is, therefore, called the assignment probability function.
It defines the soft decisions for assigning variable $x_i$ at the time instance $t$.
If a variable value $x_i$ is of a higher function value $p_i(x_i, t)$,
	then it is more likely to be assigned to the $i$-th variable  
	than any other value of a lower function value.

The assignment function $\Psi_i(x_i, t)$ is also called the assignment state function,
	representing the state of agent $i$ at the time instance $t$.
From (\ref{compute_assignment_probabilty}) we can see that
	the assignment probability function $p_i(x_i, t)$ is defined 
	as the assignment state function $\Psi_i(x_i)$ to the power $\alpha$ with normalization.

%To show the relationship, the assignment probability function $p_i(x_i, t)$ is also expressed as $\left({\bar \Psi}_i(x_i,t)\right)^{\alpha}$ 
%	in the following discussions with the bar standing for the normalization.
	
With the bar notation for normalization introduced in the subsection~\ref{proof-theorem2}, 
	the iterative update function~(\ref{cooperative_optimization_general3}) can be rewritten as
\begin{equation}
\Psi_i (x_i, t) = \sum_{\sim x_i} \left( e^{E_i(x)/\hbar} \prod_{j \not= i} \left({\bar \Psi}_j (x_j, t-1)\right)^{\alpha} \right),\quad \mbox{for $i=1,2,\ldots,n$}. 
\label{cooperative_optimization_general3b}
\end{equation}

Without loss of generality, let the utility function $u_i(x)$ for the agent $i$ be
\[ u_i (x) = e^{E_i(x)/\hbar} \ . \]
In this case, the agent $i$ tries to maximize the utility function $u_i(x)$ instead of 
	maximizing the objective function $E_i(x)$ where the former task is fully equivalent to the latter.
Accordingly, the simple form~(\ref{cooperative_optimization_general3b}) of cooperative optimization becomes
	exactly same as the iterative update function (\ref{cooperative_optimization_general3a})
	for the action payoff function $\Psi_i(x_i, t)$.
The assignment probability function $p_i(x_i, t)$ of agent $i$ in (\ref{cooperative_optimization_general3b})
	is called the strategy of player $i$ in (\ref{cooperative_optimization_general3a}).

%-----------------------------------------------------------------------------------------
\subsubsection{Some Computational Properties of Cooperative Optimization}
%-----------------------------------------------------------------------------------------

In the simple form~(\ref{cooperative_optimization_general3b}) of cooperative optimization,
	we can replace the constant $\alpha$ by $\lambda (t) w_{ij}$, where both $\lambda(t)$ and $w_{ij}$ are parameters, i.e., 
\begin{equation}
\Psi_i (x_i, t) = \sum_{\sim x_i} \left( e^{E_i(x)/\hbar} \prod_{j \not= i} \left({\bar \Psi}_j (x_j, t-1)\right)^{\lambda (t) w_{ij}} \right) \ . 
\label{cooperative_optimization_general3d}
\end{equation}

Note that a summation operator can be approximated by a maximization operator as follows:
\[ \max_{x} e^{f(x)/\hbar} \approx \sum_{x} e^{f(x)/\hbar} \ . \]
(Under the assumption that the function $f(x)$ has a unique global maximum.)

Such an approximation becomes accurate when $\hbar \rightarrow 0^+$, i.e.,
\[ \lim_{\hbar \rightarrow 0^+} \left( \max_{x} e^{f(x)/\hbar} - \sum_{x} e^{f(x)/\hbar} \right) = 0 \ . \]

With this approximation, the iterative update function~(\ref{cooperative_optimization_general3d}) becomes
\[ \Psi_i (x_i, t) = \max_{\sim x_i} \left( e^{E_i(x)/\hbar} \prod_{j \not= i} \left({\bar \Psi}_j (x_j, t-1)\right)^{\lambda (t) w_{ij}} \right) \ . \]
Taking the logarithm of the both sides, we have 
\begin{equation}
\Psi_i (x_i, t) = \max_{\sim x_i} \left( E_i(x) + \lambda (t) \sum_{j \not= i} w_{ij} \Psi_j (x_j, t-1) \right) \ . 
\label{cooperative_optimization}
\end{equation}
This is the original general form of cooperative optimization.

In this form, each agent optimizes an objective function defined at the right side of the above equation.
It is called the compromised objective function in the sense 
	that it is the linear combination of the original objective function $E_i(x)$
	for agent $i$ and the assignment state functions $\Psi_j (x_j, t-1)$ of other agents $j$ at the previous time instance $t-1$.
Given a variable value $x_i$, the function value $\Psi_i (x_i, t)$ stores 
	the maximal value of the compromised objective function with the $i$-th variable fixed to the value.

Let $\tilde{x}_i(t)$ be the value of $x_i$ with the highest function value $\Psi_i(x_i, t)$, i.e.,
\begin{equation}
\tilde{x}_i(t) = \arg \max_{x_i} \Psi_i(x_i, t) \ . 
\label{best_assignment}
\end{equation}
That value represents the best value of $x_i$ at iteration time instance $t$ 
	for maximizing the compromised objective function defined at the right side of (\ref{cooperative_optimization}).
The solution of the system at iteration time instance $t$ is the collection of those best values as follows
\[ (\tilde{x}_1(t), \tilde{x}_2(t), \ldots, \tilde{x}_n(t)), \quad \mbox{simply $\tilde{x}(t)$} \ . \]

All of the parameters $w_{ij}$s together form a $n \times n$ matrix called the propagation matrix $W$. 
To have $\sum_i E_i(x)$ as the global utility function to be maximized,
	it is required that the propagation matrix $W=(w_{ij})_{n \times n}$ is 
	non-negative, irreducible, aperiodic, and satisfying
\[ \sum^n_{i=1} w_{ij} = 1, \quad \mbox{for $j=1,2,\ldots,n$} \ . \]
	
\begin{theorem}
Given a constant cooperation strength $\lambda$ of a non-negative value less than 1 ($0 \le \lambda < 1$),
	the general form~(\ref{cooperative_optimization}) of cooperative optimization has one and only one equilibrium. 
It always converges to the unique equilibrium with an exponential rate 
	regardless of initial conditions. 
\end{theorem}

To be more general, assume that the objective function $E_i(x)$ for agent $i$ is defined on variable set $X_i$.
Recall that the solution at iteration $t$ is $\tilde{x}(t)$ (see (\ref{best_assignment})).
Let $\tilde{x}(t)(X_i)$ denote the restriction of the solution on $X_i$.

\begin{definition}
The solution $\tilde{x}(t)$ is called a consensus solution 
	if it is the optimal solution for each optimization problem defined by (\ref{cooperative_optimization}).
That is, 
\[ 
\tilde{x}(t)(X_i) = \arg \max_{X_i} \left( E_i(x) + \lambda(t) \sum_{j \not=i} w_{ij} \Psi_j(x_j, t-1)\right),~~\mbox{for $i=1,2,\ldots, n$}. \]
\end{definition}

\begin{theorem}
If the general form~(\ref{cooperative_optimization}) of cooperative optimization converges to a consensus equilibrium
	with a constant $\lambda$ satisfying $0 \le \lambda < 1$,
	then it must be the global optimum of the global objective function $E_1(x) + E_2 (x) + \cdots + E_n(x)$. 
\end{theorem}

From (\ref{cooperative_optimization}), we can see that the agents can increase the chance of reaching a consensus 
	when the value of the parameter $\lambda$ is increased.
However, when $\lambda \ge 1$, it is no longer guaranteed that any consensus equilibrium is the global optimum.
Also, the uniqueness of equilibrium is no longer guaranteed.
Assume that the maximization of $E_i(x)$, for any $i$, also leads to the maximization of the global objective function 
	$E(x)$.
Then, when $\lambda \rightarrow \infty$,
	the cooperative optimization (\ref{cooperative_optimization}) falls back to local search, 
	a classic optimization method (see Section 3.5 in \cite{Pardalos02}).
A local search algorithm can have many local optimal solutions 
	and the number of them may grow exponentially with the problem size.

In summary, the cooperative optimization algorithm~(\ref{cooperative_optimization}) is absolutely stable
	when the cooperation strength $\lambda$ is less than one ($\lambda < 1$).
Above that value, the number of equilibria may grow with the value.
As a consequence, the algorithm may become less stable because it can get stuck into one equilibrium or another.
On the other hand,
	the chance of reaching a consensus equilibrium increases.
A consensus equilibrium is guaranteed to be the global optimal one only when $\lambda < 1$.
Hence, the performance of the algorithm usually peaks
	at some positive value for the cooperation strength $\lambda$.
It deteriorates when the value is moved away from the best performing value,
	either further up or further down towards the value zero.
	
The above investigations are not on a rigorous basis.
The exact performance of the cooperative optimization algorithm~(\ref{cooperative_optimization})
	in relationship with the cooperation strength $\lambda$ is an open question.

\end{document}